\documentclass[english,twocolumn]{revtex4}
\usepackage[T1]{fontenc}
\usepackage[latin1]{inputenc}
\usepackage{graphicx}
\usepackage{times}

\makeatletter

\usepackage{babel}
\makeatother
\begin{document}

\title{Model Dependence of Lateral Distribution Functions of \\
High Energy Cosmic Ray Air Showers}

\author{Hans-Joachim Drescher, Marcus Bleicher, Sven Soff, Horst Stöcker}

\address{Institut für Theoretische Physik\\
Johann Wolfgang Goethe-Universit\"{a}t\\
 Robert-Mayer-Str 8-10. \\
 60054 Frankfurt am Main, Germany}

\begin{abstract}
The influence of high and low energy hadronic models on lateral 
distribution functions of cosmic ray air showers 
for Auger energies is explored. 
A large variety of presently used high and low energy hadron interaction 
models are analysed and the resulting lateral distribution functions 
are compared. 
We show that the slope 
% as well as the signal at 1000~m distance from the shower axis 
depends on both the high and low energy hadronic model used. 
The models are confronted with available hadron-nucleus
data from accelerator experiments.
\end{abstract}

\maketitle

\section{Introduction}

Microscopic calculations of cosmic ray air showers at energies $>10^{18}$ eV
are of crucial importance for the reconstruction of the properties
of the primary cosmic ray. Ground-array type experiments like AGASA
and Auger (working in non-hybrid mode) 
determine the energy of the primary cosmic ray by measuring 
the particle density at some distance from the shower axis. 
Muons as ground particles are interesting because they carry
 information about the composition of the primary cosmic ray; 
showers induced by heavy nuclei tend to give more muons than 
showers induced by protons or photons.

The interpretation of the observed signals depends strongly on
the quality of the air shower model and simulation.
Careful modelling of the shower development and properties is therefore 
of utmost importance. Many papers have restricted their focus 
to the dependence of air shower properties on the high energy hadronic 
model \cite{Heck:1997tw,Heck:2002yc,Heck:2000nv,Heck:2001is,Heck:2002yf,Alvarez-Muniz:2002ne}. 
This has been thought to be the main uncertainty for the determination 
of the shower maximum in the longitudinal profile. Models that 
predict a higher cross section or a higher multiplicity yield 
shower maxima at smaller slant depths, 
i.e., higher in the atmosphere (see e.g. \cite{Alvarez-Muniz:2002ne}).

In this paper we analyse the model dependence for lateral distribution
functions (LDF) for muons, electrons/positrons and photons. 
For these observables,
the low energy hadronic model becomes important, especially for the
LDF of muons, because these are decay 
products of low energy charged mesons \cite{Engel99,Engel01}.
Also the tails of LDFs of electrons/positrons and photons are
influenced by the low interaction energy region as discussed 
in \cite{Drescher:2002vp}.

\section{Air shower modelling}

For the present investigation of air shower models the SENECA framework 
is used. For an introduction and details about the SENECA model, the reader 
is referred to  \cite{Drescher:2002cr,Bossard:2000jh}.
In this approach, the intermediate energy region is calculated by solving
a set of transport equations numerically. 
The high energy models that we compare are QGSJET01 
\cite{QSJET} and SIBYLL2.1 \cite{Fletcher:1994bd,Engel:1999db};
the choices for the low energy model are GHEISHA \cite{GHEISHA},
\mbox{G-FLUKA}\cite{FLUKA} (the notation G- denotes the older version of
FLUKA as found in Geant3.21), and UrQMD 1.2.1 \cite{Bleicher:1999xi,Bass:1998ca,Agostinelli:2002hh}.
Electromagnetic showering is done by the EGS4-code \cite{EGS4}. 

UrQMD is a microscopic transport approach based on the covariant propagation
of constituent quarks and di-quarks accompanied by mesonic and baryonic
degrees of freedom. The leading hadrons of the fragmenting strings
contain the valence-quarks of the original excited hadron. 
The elementary hadronic interactions are modelled according
to measured cross sections and angular distributions. If the cross
sections are not experimentally known, detailed balance is employed
in the energy range of resonances. The partial and total decay widths
are taken from the Particle Data Group \cite{Caso:1998tx}.

Since UrQMD does not have an intrinsic cross section calculation,
it is implemented in the following way. For collisions with e.g. nitrogen
one assumes a disk with radius 5~fm, corresponding to a geometric 
cross section of 
750~mb. The projectile is then propagated over a distance which is 
chosen randomly
according to the mean free path, and is placed randomly on this disk.
The fact that the cross section was assumed too large is automatically 
compensated by a certain number of non-interactions. 
The fraction of interactions
times 750~mb gives the inelastic cross section. In a certain way,
one can picture this as nature does it: The particle passes through
the air, and whenever the transverse distance to a nucleus is too
large, it continues to propagate without interaction.

\begin{figure}
\includegraphics[%
  width=0.80\columnwidth]{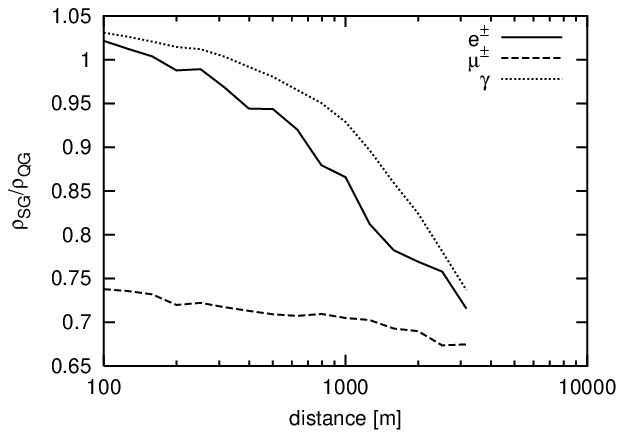}

\includegraphics[%
  width=0.80\columnwidth]{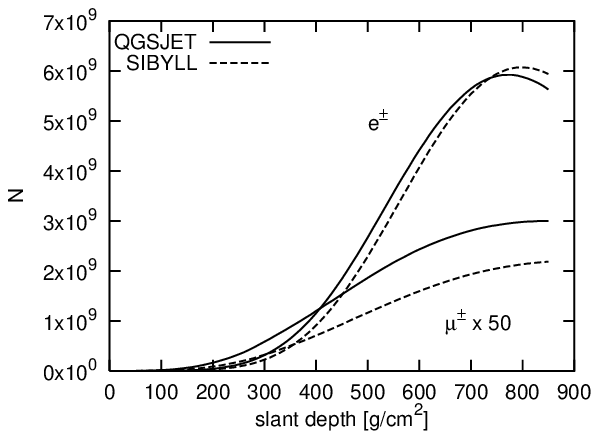}

\caption{\label{cap:QGSG} 
Differences due to high energy interaction models for $1\times10^{19}$~eV proton induced showers.
Top panel: Ratio of LDFs of the model 
combinations (SIBYLL+GEISHA)/(QGSJET+GEISHA).
Bottom panel: the longitudinal profiles (electrons/positrons and muons)
of the same combinations.}
\end{figure}

\section{Results}

\subsection{Modification of lateral distribution functions}

Lateral distribution functions are composed of three particle types,
namely
electrons/positrons, photons and muons. The former two are initiated
by the decays of high and low energetic $\pi^{0}$s, whereas the muons
are decay products of low energetic charged mesons. As pointed out in
\cite{Drescher:2002vp}, high energy pion decays determine the electromagnetic
LDF near the core. However, at large distances from the core, the contribution from
low energy pions gains importance. Muons are in general
produced by charged mesons that decay at intermediate altitudes ($\sim 5$~km). 

Let us start with a comparison of the LDFs for different high energy 
hadronic interaction  models. We compare vertical proton induced 
$1\times10^{19}$~eV showers (averaged), observed at 860~g/cm$^2$ altitude.
The kinetic energy cutoffs are $1$~MeV for electrons/photons, 
and $50$~MeV for muons. 
As previously shown with CORSIKA simulations in \cite{Heck:2002yf}, 
the present study also
gives fewer muons and more electrons for SIBYLL than for QGSJET
at small distances.
At large distances both the electromagnetic and
the muonic component are smaller in SIBYLL, 
as shown in Fig.~\ref{cap:QGSG} (top). 
This is confirmed in the average longitudinal
profile from both models, as shown in Fig.~\ref{cap:QGSG} (bottom): 
QGSJET develops higher in the atmosphere. 
The significant difference in muons can be traced back to the higher average 
multiplicity of QGSJET. This difference 
is about 25\% at 100~GeV and reaches a factor of two at $10^{20}$~eV 
\cite{Alvarez-Muniz:2002ne}. 
At low energies this is due to different implementations of parton 
distribution functions (PDFs), 
QGSJET implements a flatter choice, even though contradictory to 
A flater PDF has a higher parton density a intermediate 
$x_F$ (Feynman-x: longitudinal momentum fraction) or energy,
and higher density at lower $x_F$ or higher energy.
HERA data (see \cite{Ostapchenko:2003sj,Knapp28Icrc} for discussion). 
This leads to a somewhat higher multiplicity at low energies (ca. 100~GeV)
but helps to control particle multiplicities up to collider energies. 
For higher energies particle production increases again strongly 
since QGSJET misses higher order corrections.
Current efforts focus on resolving this problem  by introducing 
enhanced Pomeron diagrams \cite{Ostapchenko27ICRC}.
SIBYLL mimics this effect by the implementation of 
an ad-hoc energy dependent $p_\perp$ cutoff for hard processes 
\cite{Engel:1999db}. Since the main contribution of particle
production comes from hard processes, an increase of  $p_\perp$
with energy leads to a moderate rise of the multiplicity.

\begin{figure}
\includegraphics[%
  width=0.80\columnwidth]{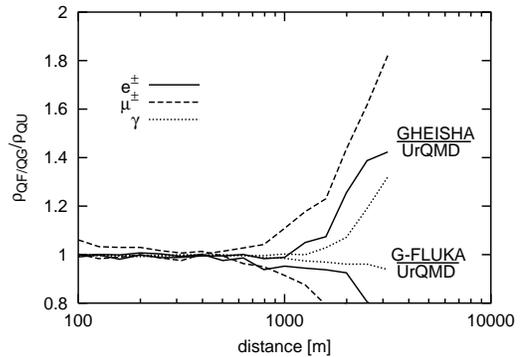}
\caption{\label{cap:ratio-QGQFQU} Dependence of LDFs for QGSJET
as high energy model, and GHEISHA, \mbox{G-FLUKA}, and UrQMD as low energy
models. }
\end{figure}

In the following, vertical $5\times10^{19}$~eV proton
induced showers are analysed. Fig.~\ref{cap:ratio-QGQFQU} shows the ratio of LDFs
for the following combination of hadronic models. QGSJET and GHEISHA
(QG), QGSJET and \mbox{G-FLUKA} (QF), and QGSJET and UrQMD (QU). The functions
are plotted as a ratio to QGSJET+UrQMD. Whereas the QGSJET+\mbox{G-FLUKA} 
and QGSJET+UrQMD models show rather
similar LDFs, resulting in a ratio close to unity, 
QGSJET+GHEISHA has a notably flatter LDF. 
The effect is most significant
for muons, somewhat less for electrons and photons. This can be understood
by the fact that most of the last collisions for muons producing hadrons 
are in the low-energy regime. The same is true for electrons and photons
at larger distances from the shower-axis. The muon ratio reaches almost
a factor of 2 at 3000 meters. 

\begin{figure}
\includegraphics[%
  width=0.80\columnwidth]{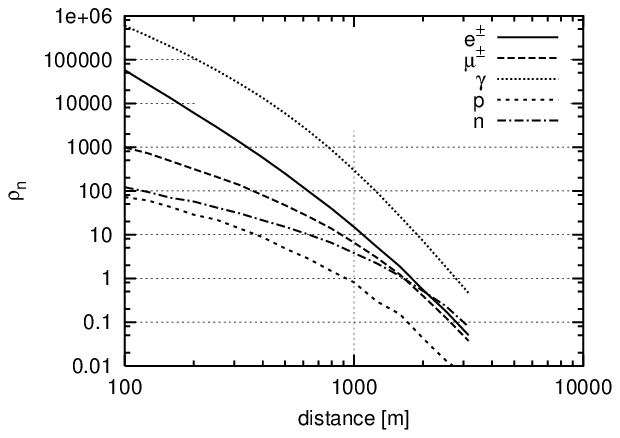}

\includegraphics[%
  width=0.80\columnwidth]{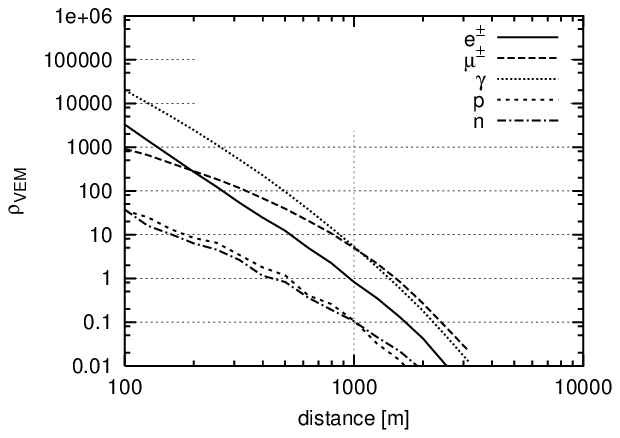}

\caption{\label{cap:LDF-and-VEM}Lateral distribution functions 
for different particle species in number densities 
and vertical equivalent muon  units for an incoming proton of 
$5\times10^{19}\textrm{eV}$.}
\end{figure}

It is especially interesting to study the 
experimental response of the different
components. At the Pierre Auger Observatory,  the density is measured 
in vertical equivalent muon units (VEM). 
A vertical minimum ionising muon counts as one particle. Since
the detectors are water tanks, muons penetrate much deeper than electrons
and photons, and are expected to give a more pronounced signal. We calculate
the detector response in terms of Cherenkov photons. A detector simulation
based on the WTANK program \cite{wtank}
has been done for the setup of an Auger water tank, by counting the
number of Cherenkov photons arriving at the positions of the photo-multipliers.
Tables with the results have been made for different incident angles, energies
and particle types and are then applied to the simulations. 
The results for an average vertical $5\times10^{19}\textrm{eV}$ shower are
shown in Fig.~\ref{cap:LDF-and-VEM}. The model combination used 
here is QGSJET01+UrQMD. One sees that the curves for muons in 
number density and VEM units almost agree, whereas photons and electrons give
less contribution per particle. Photons are still dominant due to
their high number. These results are in good agreement with 
calculations presented in the Auger Design Report \cite{unknown:1996re}. 
For completeness we show that
protons and neutrons have a negligible contribution to the total signal.

\begin{figure}
\includegraphics[width=0.80\columnwidth]{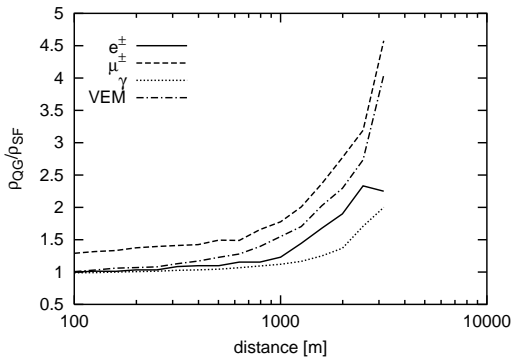}
\caption{\label{cap:ratio-LDF-and-VEM}
LDF ratios of $5\times10^{19}\textrm{eV}$ proton induced vertical showers
 for QGSJET01+GHEISHA and SIBYLL+\mbox{G-FLUKA}, the model 
combinations with the largest differences. The VEM of the 
total signal is also shown.
}
\end{figure}

The ratio of QGSJET+GHEISHA to SIBYLL+ \mbox{G-FLUKA}, together
with the VEM yields, are depicted in Fig.~\ref{cap:ratio-LDF-and-VEM}.
These model combinations give the most different results.
The effect on the slope of the VEM-LDF is even bigger because at small
distances photons dominate the signal, whereas at large distances 
the muons give the most contribution.
The sensitivity of Auger to models is somewhat higher 
than for experiments using thin plastic scintillators. 
The main reason is that water tanks are more sensitive to muons, 
since their mean path length through the material is much longer.
Further, muons become more dominant at large core distances, 
due to the flatter LDF. These two facts are important when 
choosing the density at a given distance as energy estimator.
However, as shown below, GHEISHA as low energy hadronic model can be excluded 
from accelerator data and the differences between the remaining two 
low energy models 
are not so large. With a well measured LDF and a cross-calibration 
with the fluorescence method, Auger might be able to 
discriminate between high energy hadronic models.

\begin{figure}
\includegraphics[width=0.80\columnwidth]{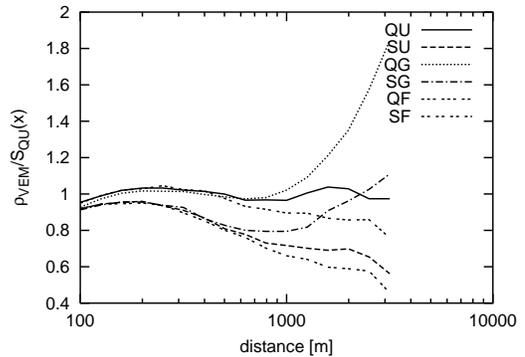}
\caption{\label{cap:VEM_all}
The LDFs in VEM units, scaled to the fitted function for QGSJET01+UrQMD. 
}
\end{figure}

The LDFs for the different models have been fitted to the following function
\begin{equation}
S(r)=C r^{-\eta}(1+(r/r_0)^2)^{-\alpha} \label{for:LDF}~,
\end{equation}
with $C$, $\eta$, $r_0$, $\alpha$ being fit parameters and $r$ being
the distance to the shower axis in meter. This function
is inspired from the one used by AGASA \cite{Takeda:2002at}. 
The resulting parameters are shown in Table \ref{table}.
Fig.~\ref{cap:VEM_all} shows the LDFs in VEM units for all 
model combinations, scaled by function (\ref{for:LDF}), using 
the parameters for QGSJET+UrQMD. 
One sees how at small distances, the high energy model determines the 
shape; at larger distances deviations result from differences in the 
low energy hadronic models.

%QU:40015547062.3239 3.09753565377978 1539.86832999103 1.62861029124084
%SU:48783271856.3648 3.14791206600281 1136.06676828536 1.29743234240298
%QF:31250939509.7521 3.0463831867535  1125.98011111769 1.3436088698253
%SF:41491529654.5612 3.11474066244941 985.241591846936 1.26261960831787
%QG:31736971691.8906 3.05448901893405 986.230804504059 0.86554525345278
%SG:48810131338.7965 3.14915344216799 1036.2468725502 0.955254072476208

\begin{table}
\begin{tabular}{|c|c|c|c|c|}
\hline
Model & $C$ & $\eta$ &  $r_0 [m]$  & $\alpha$ \tabularnewline
\hline
QGSJET01+GHEISHA&3.17$\times10^{10}$ & 3.05 & 986 & 0.87\tabularnewline
\hline
QGSJET01+UrQMD&4.00 $\times10^{10}$ & 3.10 & 1540 & 1.63\tabularnewline
\hline
QGSJET01+\mbox{G-FLUKA}&3.13$\times10^{10}$  &3.05 & 1126 & 1.34\tabularnewline
\hline
SIBYLL+GHEISHA&4.88$\times10^{10}$  &3.15 & 1036 & 0.96\tabularnewline
\hline
SIBYLL+UrQMD&4.88$\times10^{10}$  &3.15 & 1136 & 1.30\tabularnewline
\hline
SIBYLL+\mbox{G-FLUKA}&4.15$\times10^{10}$ & 3.11 & 985 & 1.26\tabularnewline
\hline
\end{tabular}
\caption{\label{table}
The fit parameters for function (\ref{for:LDF}) for all model combinations.}
\end{table}

\subsection{Comparisons of models with data}

\begin{figure}
\includegraphics[%
  width=0.80\columnwidth]{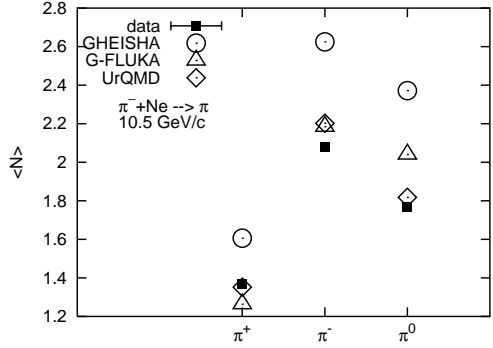}

\caption{\label{cap:piNe}Multiplicities for pions of a $\pi^{-}$+Ne reaction
at 10.5~GeV/c. Data are taken from \cite{Yeager:1977ym}.}
\end{figure}

In this section the behaviour of the different
hadronic models and their influence on the LDFs is analysed. Here we focus on
data available for hadron-nucleus collisions in the relevant energy domain. 
Light nuclei are chosen which are somewhat comparable to air. 
Fig.~\ref{cap:piNe} addresses the
mean $4\pi$ multiplicity of pions for $\pi^{-}+$Ne reactions 
at $10.5$~GeV/c \cite{Yeager:1977ym}.
One clearly observes that GHEISHA produces too many pions, 
while UrQMD and \mbox{G-FLUKA}
are in line with the measured data. 

\begin{figure}
\includegraphics[%
  width=0.80\columnwidth]{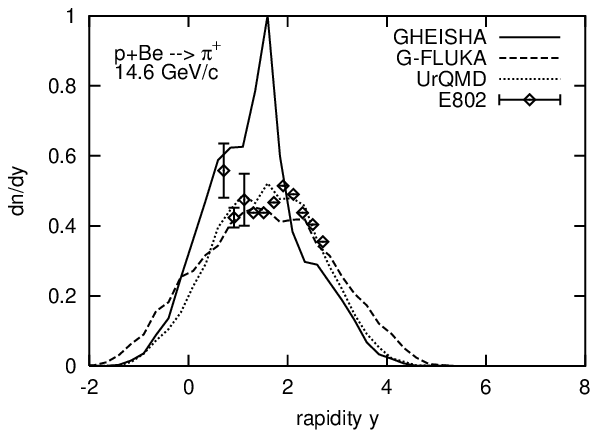}

\includegraphics[%
  width=0.80\columnwidth]{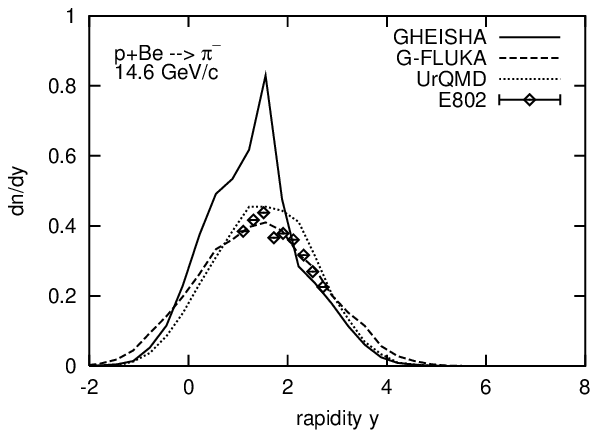}

\includegraphics[%
  width=0.80\columnwidth]{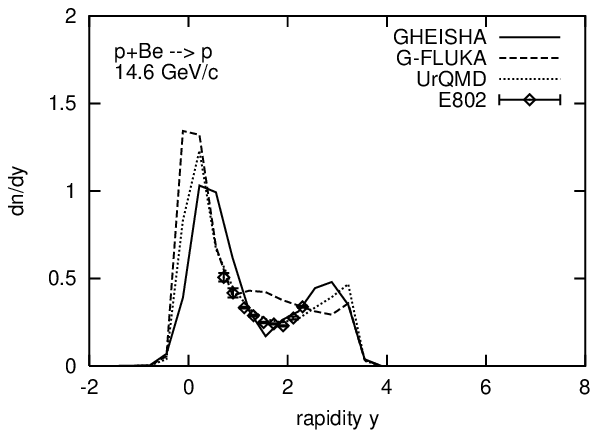}

\caption{\label{cap:pBe}Rapidity spectra of final state pions and protons
in p+Be reactions at 14.6 GeV/c. Data
are taken from E802 \cite{Abbott:1992en}.}
\end{figure}
To explore the longitudinal momentum distribution, we confront the
models with rapidity spectra for p+Be reactions at 14.6~GeV/c taken
at the AGS. The fact that GHEISHA shows a strange peak in the rapidity
spectra is well-known \cite{Ferrari1996}. In addition, GHEISHA yields too
high multiplicities for pions, and \mbox{G-FLUKA} overestimates the
longitudinal momentum loss for protons.
The stopping of the initial proton is well described by UrQMD and GHEISHA. 

\begin{figure}
\includegraphics[%
  width=0.80\columnwidth]{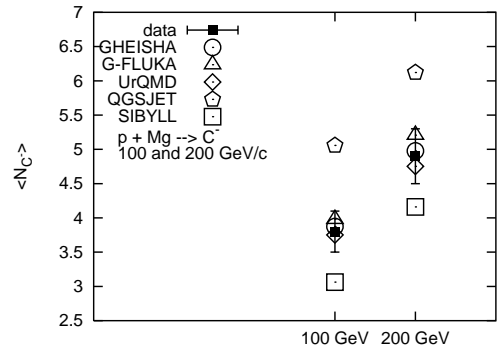}

\caption{\label{cap:pMg}Multiplicity of negatively charged particles for
a p+Mg reaction at 200 GeV/c. Data are taken from NA35 \cite{Baechler:1991jh}.}
\end{figure}
\begin{figure*}
\includegraphics[%
  width=1.0\linewidth]{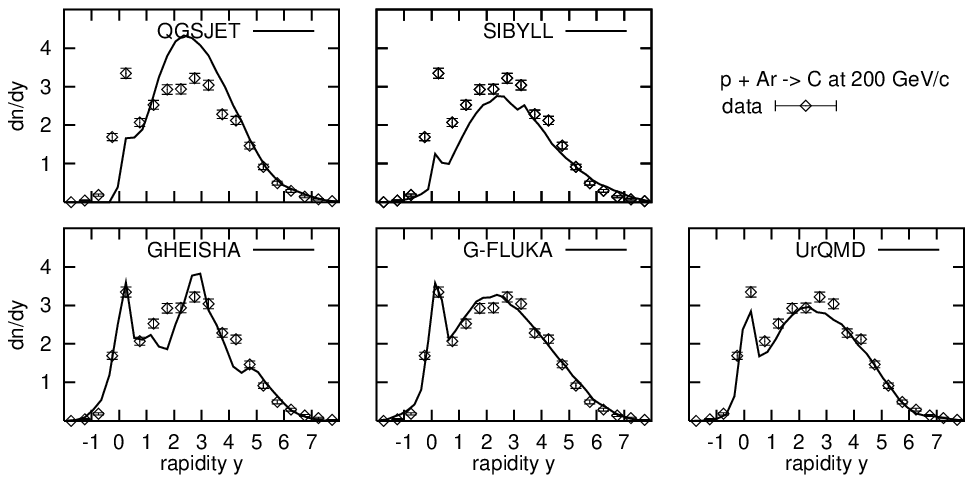}

\caption{\label{cap:pAr_pos}Rapidity distributions for the reaction 
p+Ar $\rightarrow$ charged hadrons.
Data are taken from \cite{DeMarzo:1982rh}.}
\end{figure*}
\begin{figure*}
\includegraphics[%
  width=1.0\textwidth]{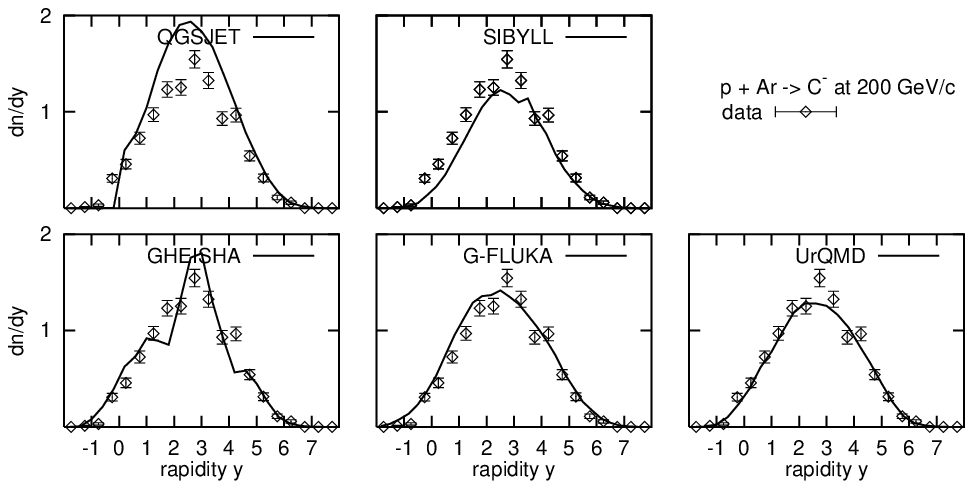}

\caption{\label{cap:pAr_neg}Rapidity distributions for the reaction 
p+Ar $\rightarrow$ negatively charged hadrons. 
Data are taken from \cite{DeMarzo:1982rh}.}
\end{figure*}
\begin{figure*}
\includegraphics[%
  width=1.0\textwidth]{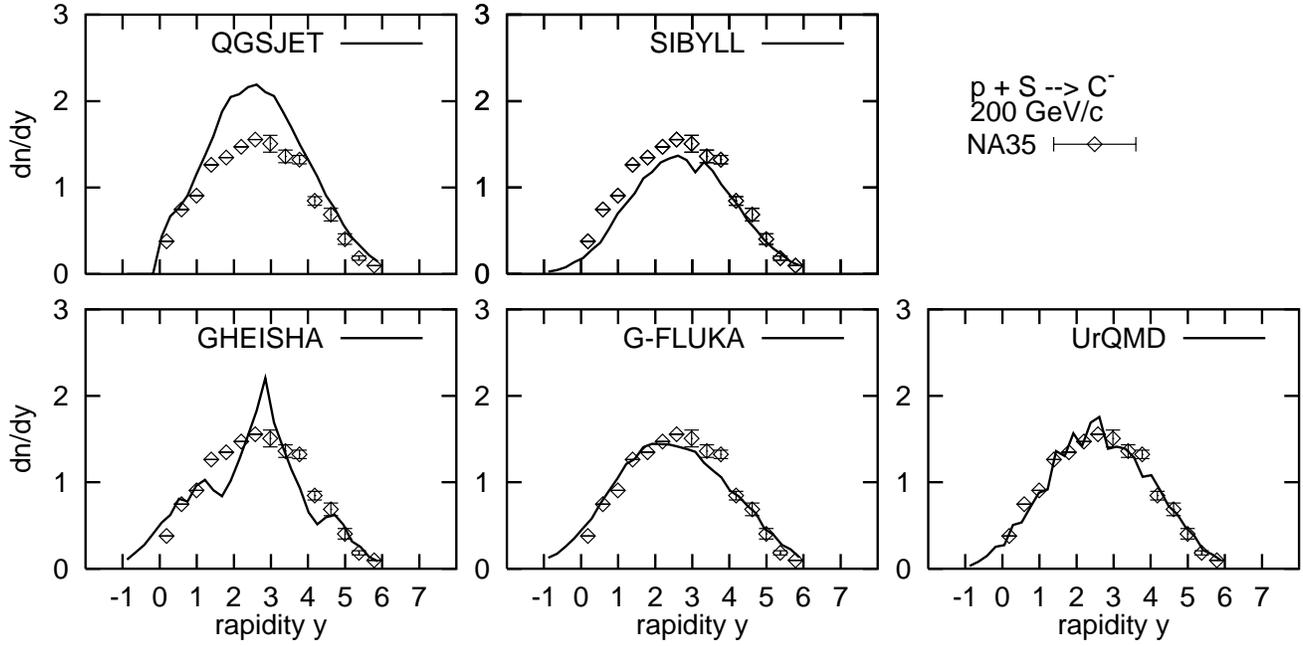}

\caption{\label{cap:pS_neg}Rapidity distributions for the reaction 
p+S $\rightarrow$ negatively charged hadrons.
Data are taken from NA35 \cite{Alber:1998sn}.}
\end{figure*}
\begin{figure*}
\includegraphics[%
  width=1.0\textwidth]{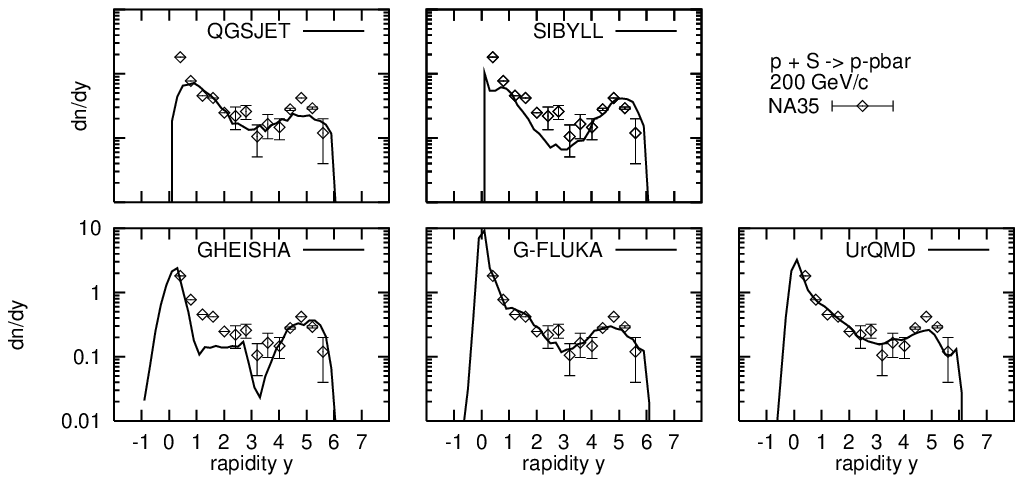}

\caption{\label{cap:pS_net}Rapidity distributions for the reaction 
p+S $\rightarrow$ net protons.
Data are taken from NA35 \cite{Alber:1998sn}.}
\end{figure*}

At higher energies, we can add QGSJET and SIBYLL to the comparison. 
These models are commonly employed at energies larger
than $E_{\rm beam}\sim$~100~GeV. For example, CORSIKA uses 80~GeV as 
transition value. Fig.~\ref{cap:pMg}
shows that QGSJET overestimates and SIBYLL underestimates
the negatively charged hadron multiplicities in p+Mg 
reactions at 100 and 200~GeV/c. 
The three other model results are in line with the data (NA35 \cite{Baechler:1991jh}). 

\clearpage

Figs.~\ref{cap:pAr_pos} and \ref{cap:pAr_neg} show rapidity spectra
of (negatively) charged hadrons 
for p+Ar reactions \cite{DeMarzo:1982rh}. Again, QGSJET and SIBYLL do deviate from the data. 
GHEISHA still produces the peak, which is located such that the total
multiplicity is well described. \mbox{G-FLUKA} and UrQMD give good descriptions
of the data. A similar behaviour is obtained for  p+S collisions at 200~GeV/c 
(NA35 data \cite{Alber:1998sn}), as demonstrated in Figs.~\ref{cap:pS_neg} and \ref{cap:pS_net}. 
Note that the  analysis has been
performed with the experimental trigger demanding a minimum of 5 charged
particles being produced in the reaction.
 
As above, the longitudinal momentum loss of the protons 
(stopping) is best described by \mbox{G-FLUKA} and UrQMD. 

\begin{figure}
\includegraphics[%
  width=1.0\columnwidth]{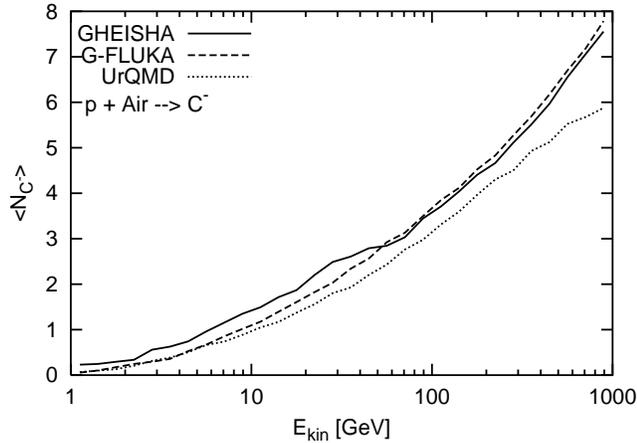}

\caption{\label{cap:meanL}Mean multiplicities of negatively charged hadrons 
for the discussed set of low energy models.}
\end{figure}
\begin{figure}
\includegraphics[%
  width=1.0\columnwidth]{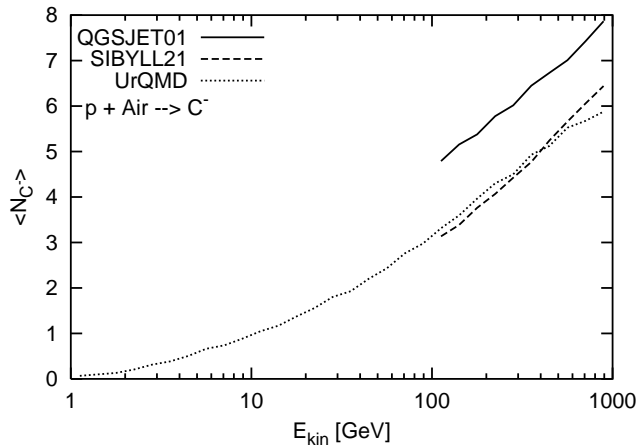}

\caption{\label{cap:meanH}Mean multiplicities of negatively charged hadrons 
for the discussed set of high energy models. UrQMD results are given as
a comparison.}
\end{figure}
An overall comparison is given in Figs.~\ref{cap:meanL} and \ref{cap:meanH},
where the mean multiplicity of negatively charged hadrons 
is plotted as a function of the incident kinetic energy.
Below 100~GeV GHEISHA produces more particles per collision; this
leads to the higher muon densities at large distances. In the same
way the multiplicity of produced $\pi^{0}$s changes the shape of
the LDF of electrons and photons, only at larger distances, where
this energy region becomes more dominant. 
UrQMD is slightly lower in multiplicity than \mbox{G-FLUKA}. 
The fact that the latter results in smaller signals at 
large distances might be due to differences in the total
cross section, which is neglected in this analysis.
The overestimation of particle production by QGSJET in the transition 
region above 100~GeV is significant.  
Since QGSJET is the most used model in air 
shower simulations, these deviations do deserve further studies,
because the energy region above 100~GeV is certainly important 
for muon production. 

\section{Conclusions}

\begin{itemize}

\item We have demonstrated that the LDFs for high energy air showers
depend on the high and also
on the low energy hadronic models. The tails are strongly influenced by the
low energy model, which therefore modifies the slope of the LDFs. 

\item GHEISHA, which is commonly used as low energy hadronic model, 
 reproduces less accurately available data of hadron induced interactions. 
\mbox{G-FLUKA} and UrQMD describe the data better. 

\item In the transition region around 100~GeV, QGSJET predicts too high 
particle multiplicities, whereas SIBYLL is somewhat low. 
The present findings suggest to increase the transition energy. 
UrQMD is tested up 
to RHIC energies (21~TeV), and could be employed up to this value.

\end{itemize}
\begin{acknowledgments}
This work has been supported by the German BMBF/DESY.
The computational resources have been provided by the 
Center for Scientific Computing in
Frankfurt am Main.
\end{acknowledgments}
\bibliographystyle{unsrt}
\bibliography{ref}

\begin{thebibliography}{10}

\bibitem{Heck:1997tw}
D.~Heck, G.~Schatz, and J.~Knapp.
\newblock {\em Nucl. Phys. B (Proc. Suppl.)}, 52:139, 1997.

\bibitem{Heck:2002yc}
D.~Heck et~al.
\newblock In {\em {Proceedings of the 27th International Cosmic Ray
  Conference}}, page 233, Hamburg, Aug. 7-15 2001.

\bibitem{Heck:2000nv}
D.~Heck.
\newblock In T.~Csorgo, editor, {\em Proc. XXX Int. Symp. Multiparticle
  Dynamics}, page 252, Tihany (Hungary), 9-15 Oct. 2000. World Scientific
  (Singapore) 2001.

\bibitem{Heck:2001is}
D.~Heck.
\newblock {\em Nachr. Forsch. Zentr. Karlsruhe}, 33:113, 2001.

\bibitem{Heck:2002yf}
D.~Heck et~al.
\newblock {\em Nucl. Phys. B (Proc. Suppl.)}, 122:364, 2002.

\bibitem{Alvarez-Muniz:2002ne}
J.~Alvarez-Muniz et~al.
\newblock {\em Phys. Rev.}, D66:033011, 2002.

\bibitem{Engel99}
R.~Engel et~al.
\newblock In I.~Sarcevi and C.-I. Tan, editors, {\em Providence 1999, QCD and
  multiparticle production}, page 457, Providence, Rhode Island, Aug. 9-13
  1999. World Scientific (Singapore).

\bibitem{Engel01}
R.~Engel.
\newblock In {\em {Proceedings of the 27th International Cosmic Ray
  Conference}}, page 181, Hamburg, Aug. 7-15 2001.
\newblock (rapporteur talk).

\bibitem{Drescher:2002vp}
H.J. Drescher and G.R. Farrar.
\newblock {\em Astropart. Phys.}, 19:235, 2003.

\bibitem{Drescher:2002cr}
H.J. Drescher and G.R. Farrar.
\newblock {\em Phys. Rev.}, D67:116001, 2003.

\bibitem{Bossard:2000jh}
G.~Bossard et~al.
\newblock {\em Phys. Rev.}, D63:054030, 2001.

\bibitem{QSJET}
N.N. Kalmykov, S.S. Ostapchenko, and A.I. Pavlov.
\newblock {\em Nucl. Phys. B (Proc. Suppl.)}, 52:17, 1997.

\bibitem{Fletcher:1994bd}
R.~S. Fletcher et~al.
\newblock {\em Phys. Rev.}, D50:5710, 1994.

\bibitem{Engel:1999db}
R.~Engel et~al.
\newblock In {\em Proceedings of the 26th International Cosmic Ray Conference},
  page 415, Salt Lake City, Utah, Aug 17-25 1999.

\bibitem{GHEISHA}
H.~Fesefeldt.
\newblock {\em PITHA 85/02, RWTH Aachen}, 1985.

\bibitem{FLUKA}
A.~Fasso et~al.
\newblock In {\em Proceedings of the Workshop on Simulating Accelerator
  Radiation Environments}, page 134. Los Alamos Report LA-12835-C, 1992.

\bibitem{Bleicher:1999xi}
M.~Bleicher et~al.
\newblock {\em J. Phys.}, G25:1859, 1999.

\bibitem{Bass:1998ca}
S.~A. Bass et~al.
\newblock {\em Prog. Part. Nucl. Phys.}, 41:225, 1998.

\bibitem{Agostinelli:2002hh}
S.~Agostinelli et~al.
\newblock {\em Nucl. Instrum. Meth.}, A506:250, 2003.

\bibitem{EGS4}
W.~R. Nelson et~al.
\newblock SLAC-265, Stanford Linear Accelerator Center, 1985.

\bibitem{Caso:1998tx}
C.~Caso et~al.
\newblock {\em Eur. Phys. J.}, C3:1, 1998.

\bibitem{Ostapchenko:2003sj}
S.~S. Ostapchenko.
\newblock {\em J. Phys.}, G29:831, 2003.

\bibitem{Knapp28Icrc}
M.~Zha et~al.
\newblock In {\em Proceedings of the 28th International Cosmic Ray Conference},
  page 515, Tsukuba, Jul. 31-Aug. 7 2003.

\bibitem{Ostapchenko27ICRC}
S.~Ostapchenko et~al.
\newblock In {\em Proceedings of the 27th International Cosmic Ray Conference},
  page 446, Hamburg, Aug. 7-15 2001.

\bibitem{wtank}
J.R.T de~Mello~Neto.
\newblock {\em Auger technical notes: GAP-1998-20}.
\newblock http://www.auger.org.

\bibitem{unknown:1996re}
{http://www.auger.org}.

\bibitem{Takeda:2002at}
M.~Takeda et~al.
\newblock {\em Astropart. Phys.}, 19:447, 2003.

\bibitem{Yeager:1977ym}
W.~M. Yeager et~al.
\newblock {\em Phys. Rev.}, D16:1294, 1977.

\bibitem{Abbott:1992en}
T.~Abbott et~al.
\newblock {\em Phys. Rev.}, D45:3906, 1992.

\bibitem{Ferrari1996}
A.~Ferrari and P.R. Sala.
\newblock {\em Atlas int. Note PHYS-No-086 (CERN)}, 1996.

\bibitem{Baechler:1991jh}
J.~Baechler et~al.
\newblock {\em Z. Phys.}, C51:157, 1991.

\bibitem{DeMarzo:1982rh}
C.~De~Marzo et~al.
\newblock {\em Phys. Rev.}, D26:1019, 1982.

\bibitem{Alber:1998sn}
T.~Alber et~al.
\newblock {\em Eur. Phys. J.}, C2:643, 1998.

\end{thebibliography}
\end{document}